\documentclass[copyright]{eptcs}
\providecommand{\event}{T. Neary, D. Woods, A.K. Seda and N. Murphy (Eds.):\\ The Complexity of Simple Programs 2008.}
\providecommand{\volume}{1}
\providecommand{\anno}{2009}
\providecommand{\firstpage}{235}
\usepackage{breakurl}

\usepackage{graphicx,amsmath,amsfonts,amssymb}

\def\NN{\mathbb{N}}
\newcommand\R{R}
\newtheorem{thm}{Theorem}

\def\ie{i.e.}

\title{New Choice for Small Universal Devices: Symport/Antiport P Systems}
\author{Sergey Verlan
    \institute{LACL, D\'epartement Informatique, Universit\'e Paris Est,\\
        61 av. G\'en\'eral de Gaulle, 94010 Cr\'eteil, France}
    \institute{Institute of Mathematics and Computer Science\\
        Academy of Sciences of Moldova, Academiei, 5, MD-2028, Moldova}
    \email{verlan@univ-paris12.fr}
    \and
    Yurii Rogozhin
    \institute{Institute of Mathematics and Computer Science\\
    Academy of Sciences of Moldova, Academiei, 5, MD-2028, Moldova}
    \institute{Rovira i Virgili University,\\
    Research Group on Mathematical Linguistics,\\
    Pl. Imperial T\`arraco 1, 43005 Tarragona, Spain}
    \email{rogozhin@math.md}
}

\def\titlerunning{New Choice for Small Universal Devices: Symport/Antiport P Systems}
\def\authorrunning{S. Verlan and Y. Rogozhin}

\begin{document}
%\begin{frontmatter}
\maketitle
\begin{abstract}
Symport/antiport P systems provide a very simple machinery inspired by
corresponding operations in the living cell. It turns out that systems of small
descriptional complexity are needed to achieve the universality by these
systems. This makes them a good candidate for small universal devices replacing
register machines for different simulations, especially when a simulating
parallel machinery is involved. This article contains survey of these systems
and presents different trade-offs between parameters.
\end{abstract}
%\begin{keyword}
%P~systems \sep Symport\sep Antiport\sep Multiset rewriting\sep Universal
%computations
%\end{keyword}
%\end{frontmatter}

\section{Introduction}

The idea of symport/antiport P systems comes from simple observations in cell
biology. In a living cell, there is a permanent chemical exchange with the
environment. Water, ions and other chemicals enter or exit the cell depending
on its necessity. Some of these exchanges use a \emph{passive transport} where
no energy is consumed and the chemicals are moved along the chemical gradient,
while others use an \emph{active transport}, which consumes energy in order to
move chemicals against the gradient. Very often the active transport uses
\emph{co-transporters}, \ie{} molecules that facilitate the penetration of the
transported substance through the cell membrane. The most common
co-transporters either travel together with the transported substance, in this
case we speak about \emph{symport}, or they are exchanged with the transported
substance, in this case we speak about \emph{antiport}.

This transport mechanism is formalized by symport/antiport P
systems~\cite{Pauns02}, \cite{Pbook} which abstract the cell by a set of nested
compartments enclosed by membranes and chemicals by a multiset of objects. The
symport transport is then represented by a rule $(y,out)$ or $(y,in)$ that
specifies that objects present in the multiset $y$ travel together outside or
inside the current compartment. The antiport is formalized by the rule
$(x,out;y,in)$ which indicates that objects given by $x$ and present in the
compartment will exchange with objects given by $y$ situated outside the
compartment.

The evolution of a symport/antiport P system is done in a maximally parallel
way (other evolution strategies are discussed in~\cite{FreundVerlan07}),
starting from an initial distribution of objects in membranes and the result is
obtained by counting objects in some membrane when the system cannot evolve
anymore.

Further generalization of the model leads to symport/antiport \emph{tissue} P
systems where the underlying membrane structure is no more represented by a
tree as in the case of P systems but by an arbitrary graph corresponding to a
tissue of cells. More generalizations and a presentation of P systems (not
necessarily using symport and antiport operations) can be found in~\cite{Pbook}
and~\cite{Ppage}.

The computational model given by symport/antiport (tissue) P systems is very
simple, however it was shown that if a cooperation of three objects is
permitted, then one membrane is sufficient to generate all recursively
enumerable sets of numbers~\cite{FreundPaun02} and~\cite{FriscoHoogeboom02}.
After that other descriptional complexity parameters stared to be investigated,
in particular, systems with minimal symport or antiport, where only two objects
can cooperate. Such systems are of great interest because the biological
variants of symport and antiport involve only two objects in most of the cases.
These systems first were investigated in~\cite{BernardiniGheorghe03}, where
nine membranes were used to achieve computational completeness. This number was
progressively decreased and finally established to two membranes
in~\cite{AlhazovRogozhin06WMC}.

Other complexity parameters like the number of used objects or the number of
rules were investigated and trade-offs between different parameters were
established. In this article we present a survey of different complexity
measures and best known results.

\section{Definitions}

We recall here some basic notions of formal language theory we need in the rest
of the paper. We refer to \cite{hand} for further details.

We denote by $\NN$ the set of all non-negative integers. Let
$O=\{a_1,\dots{},a_k\}$ be an alphabet. A \emph{finite} \emph{multiset} $M$
over $O$ is a mapping $M:O\longrightarrow \NN$, \ie{}, for each $a\in O$,
$M(a)$ specifies the number of occurrences of $a$ in $M$. The size of the
multiset $M$ is $|M| =\sum_{a\in O}M(a) $. A multiset $M$ over $O$ can also be
represented by any string $x$ that contains exactly $M(a_i)$ symbols $a_i$ for
all $1\leq i\leq k$, e.g., by $a_1^{M(a_1)}\dots{}a_k^{M(a_k)}$, or else by the
set $\{a_i^{M(a_i)}\mid 1\leq i\leq k\}$. For example, the multiset over $\{
a,b,c\} $ defined by the mapping $a\rightarrow 3,b\rightarrow 1,c\rightarrow 0$
can be specified by $a^3b$ or $\{a^3,b\}$. An empty multiset is represented by
$\lambda$.

We may also consider mappings $M$ of form $M:O\longrightarrow \NN\cup
\{\infty\}$, \ie{}, elements of $M$ may have an infinite multiplicity; we shall
call  them \emph{infinite multisets}.

In the following we briefly recall the basic notions concerning P systems with
symport/antiport rules. For more details on these systems and on P systems in
general, we refer to~\cite{Pbook}.

A P system with symport/antiport of degree $n$ is a construct
$$\Pi = (O, \mu, w_1, \ldots , w_n, E, \R_1, \ldots ,\R_n, i_0), $$
 where:

\begin{enumerate}
\item  $O$ is a finite alphabet of symbols called objects,
\item $\mu$ is a membrane structure consisting of $n$ membranes that are
    labeled in a one-to-one manner by $1,2,\ldots, n$.
\item $w_i \in O^*$, for each $1 \le i \le n$ is a
%{\em finite}
multiset
%(\emph{i.e.} multiset where elements are present in finite number of copies)
of objects associated with the region $i$ (delimited by membrane $i$),
%$w_0$ is the finite multiset of objects from $V$ associated with the environment,
\item $E\subseteq O$ is the set of objects that appear in the environment
    in infinite numbers of copies,
\item $\R_i$, for each $1 \le i \le n$, is a finite set of symport/antiport
    rules associated with the region $i$ and which have the following form
    $(x,in)$, $(y, out)$, $(y,out;x,in)$,
%or $(y,in;x,out)$,
where $x,y \in O^*$,
\item $i_0$ is the label of an elementary membrane of $\mu$ that identifies
    the corresponding output region.
\end{enumerate}

A symport/antiport P system  is defined as a computational device consisting of
a set of $k$ hierarchically nested membranes that identify $k$ distinct regions
(the membrane structure $\mu$), where to each region $i$ there are assigned a
multiset of objects $w_i$ and a finite set of symport/antiport rules $\R_i$, $1
\le i \le n$. A symport rule $(x,in)\in \R_i$ permits to move $x$ into region
$i$ from the immediately outer region. Notice that rules of the form $(x,in)$,
where $x \in E^*$ are forbidden in the skin (the outermost) membrane. A symport
rule $(x,out)\in R_i$ permits to move the multiset $x$ from region $i$ to the
outer region. An antiport rule $(y, out; x, in)$ exchanges two multisets $y$
and $x$, which are situated in region $i$ and the outer region of $i$
respectively.

A computation in a symport/antiport P system is obtained by applying the rules
in a non-deterministic maximally parallel manner, i.e. all rules that can be
applied together should be applied. Other possibilities not using the maximal
parallelism are discussed in~\cite{FreundVerlan07}. The computation is
restricted to moving objects through membranes, since symport/antiport rules do
not allow the system to modify the objects placed inside the regions.
Initially, each region $i$ contains the corresponding finite multiset $w_i$;
whereas the environment contains only objects from $E$ that appear in
infinitely many copies.

A computation is successful if starting from the initial configuration it
reaches a configuration where no  rule can be applied. The result of a
successful computation is the natural number that is obtained by counting the
objects that are presented in region $i_0$. Given a P system $\Pi$, the set of
natural numbers computed in this way by $\Pi$ is denoted by $N (\Pi)$.

We denote by $NOP_n(sym_r,anti_t)$  the family of sets of natural numbers that
are generated by a P system with symport/antiport of degree at most $n > 0$,
symport rules of size at most $r \ge 0$, and antiport rules of size at most $t
\ge 0$.  The size of a symport rule $(x,in)$ or $(x,out)$ is given by $|x|$ ,
while the size of an antiport rule $(y,out;x,in)$ is given by $max\{|x|,
|y|\}$.  We denote by $NRE$ the family of recursively enumerable sets of
natural numbers.

P systems as defined above have an underlying tree-like membrane structure. It
is possible to apply a similar reasoning to an arbitrary graph. This leads us
to the idea of tissue P systems.

A \emph{tissue P system with symport/antiport} of degree $n\ge1$ is a construct

$$\Pi=(O,G,w_1,\dots{},w_n,E,R,i_0),$$

\noindent where $O$ is the alphabet of objects and  $G$ is the underlying
directed labeled graph of the system. The graph $G$ has $n+1$ nodes and the
nodes are numbered from $0$ to $n$. We shall also call nodes from $1$ to $n$
cells and node $0$ the environment. There is an edge between each cell $i$,
$1\le i\le n$, and the environment. Each cell contains a multiset of objects,
initially cell $i$, $1\le i\le n$, contains multiset $w_i$. The environment is
a special node which contains symbols from $E$ in infinite multiplicity as well
as a finite multiset over $O\setminus E$, but initially this multiset is empty.
The symbol $i_0\in (1\dots{}n)$ indicates the output cell, and $R$ is a finite
set of rules (associated to edges) of the following forms:

\begin{enumerate}
\item $(i,x,j)$, $0\le i\le n,0\le j\le n, i\ne j$, $x\in O^+$ and not
    $i=0\ \&\ x\in E^+$ (symport rules for the communication).
\item $(i,x/y,j)$, $0\le i,j\le n, i\ne j$, $x,y\in O^+$(antiport rules for
    the communication).
\end{enumerate}

We remark that $G$ may be deduced from relations of $R$. More exactly, $G$
contains $n+1$ vertices and there is an oriented edge between vertex $i$ and
$j$ if and only if there is a rule $(i,x,j)$ in $R$ and edges between $i$ and
$j$ and $j$ and $i$ if and only if there is a rule $(i,x/y,j)$ in $R$. However,
we prefer to indicate both $G$ and $R$ because it simplifies the presentation.

The rule $(i,x,j)$ sends a multiset of objects $x$ from node $i$ to node $j$.
The rule $(i,x/y,j)$ exchanges multisets $x$ and $y$ situated in nodes $i$ and
$j$ respectively. The size of symport rule $(i,x,j)$ is equal to $|x|$, while
the size of an antiport rule is equal to $|x|+|y|$.

As in the case of P systems a computational step is made by applying all
applicable rules from $R$ in a non-deterministic maximal parallel way. A
configuration of the system is an $(n+1)$-tuple $(z_0,z_1,\dots{},z_n)$ where
each $z_i,1\le i\le n$, represents the contents of cell $i$ and $z_0$
represents the multiset of objects that appear with a finite multiplicity in
the environment (initially $z_0$ is the empty multiset). The computation stops
when no rule may be applied. The result of a computation is given by the number
of objects situated in cell $i_0$, \ie{}, by the size of the multiset from cell
$i_0$.

We denote by $NOtP_n(sym_p,anti_q)$ the family of all sets of numbers computed
by tissue P systems with symport/antiport of degree at most $n$ and which have
symport rules of size at most $p$ and antiport rules of size at most $q$.

The following theorem shows the basic results for symport/antiport [tissue] P
systems:

\begin{thm}\label{thm:sa3}
$NO[t]P_1(sym_3)=NO[t]P_1(anti_3)=RE.$
\end{thm}

We can also consider accepting (tissue) P systems where an input multiset is
placed in some fixed cell/membrane and it is accepted if and only if the
corresponding system halts. Theorem~\ref{thm:sa3} holds as well in the
accepting case, however it is possible to use a deterministic construction for
the proof.

%cond uniport = 14

\section{Size of rules}

%Theorem~\ref{thm:1ma}
Theorems from the previous section show that using symport or antiport rules of
size three the computational completeness is achieved with only one membrane.
The situation changes completely if rules of size~two, called \textit{minimal
antiport} or \textit{minimal symport} rules, are considered -- in one membrane
or cell, we only get finite sets:

\begin{thm}\label{thm:1cell}
$NO[t]P_{1}(sym_1,anti_2)\cup NO[t]P_{1}(sym_2)\subseteq NFIN$.
\end{thm}

The theorem follows from the fact that the number of symbols inside the
membrane cannot be increased using minimal symport or antiport rules. Hence at
least two membranes are needed for the computational completeness. This number
is sufficient, as the following result holds.

\begin{thm}
$NO[t]P_2(sym_1,anti_2)= NO[t]P_2(sym_2)= NRE$.
\end{thm}

The proof significantly differs if tissue or tree-like P systems are
considered. In the tissue case, the proof is based on the possibility to reach
a membrane from another one by two roads, directly or via the environment,
which have a different length. In this way, a temporal de-synchronization of
pairs of objects is obtained and it can be used to simulate the instructions of
a register machine.

Moreover, in the tissue case, we have a deterministic construction for the
acceptance of recursively enumerable sets. In the tree-like case it is not
possible to use a similar technique, because only the root is connected to the
environment, which considerably restricts the accepting power of deterministic
P systems:

\begin{thm}
For any deterministic P system with minimal symport and minimal antiport rules
(of type $sym_{2}$ and $anti_{2}$), the number of objects present in the
initial configuration of the system cannot be increased during halting
computations.
\end{thm}

Hence, deterministic P systems with minimal symport and antiport rules with any
number of membranes can generate only finite languages.

However, if non-deterministic systems are considered, then it is possible to
reach computational completeness for the accepting case with two membranes: an
initial pumping phase is performed to introduce a sufficient number of working
objects needed to carry out the computation (a non-deterministic guess for the
number of working objects is done). After that, the system simulates a register
machine thereby consuming the number of working objects.

\subsection{Generalized Minimal Communication}\label{ssec:genmincomm}

We can generalize the idea of minimal antiport and symport and introduce the
concept of\textit{\ minimal interaction tissue P systems}. These are tissue P
systems where at most two objects may interact, i.e., one object is moved with
respect to another one. Such interactions can be described by rules of the form
$( a,i) ( b,j) \rightarrow ( a,k)( b,l) $, which indicate that if symbol $a$ is
present in membrane $i$ and symbol $b$ is present in membrane $j$, then $a$
will move to membrane $k$ and $b$ will move to membrane $l$. We may impose
several restrictions on these interaction rules, namely by superposing several
cells. Some of these restrictions directly correspond to antiport or symport
rules of size $2$.

Below we define all possible restrictions (modulo symmetry): let $O$ be an
alphabet and let $( a,i) ( b,j) \rightarrow (
a,k) ( b,l) $ be an interaction rule with $a,b\in O$, $%
i,j,k,l\geq 0$. Then we distinguish the following cases:

\begin{enumerate}\parskip -1pt
\item $i=j=k\neq l$: the \textit{conditional-uniport-out rule} sends $b$ to
    membrane $l$ provided that $a$ and $b$ are in membrane $i$.

\item $i=k=l\neq j$: the \textit{conditional-uniport-in rule} brings $b$ to
    membrane $i$ provided that $a$ is in that membrane.

\item $i=j\neq k=l$: the \textit{symport2 rule} corresponds to the minimal
    symport rule, i.e., $a$ and $b$ move together from membrane $i$ to $k$.

\item $i=l\neq j=k$: the \textit{antiport1 rule} corresponds to the minimal
    antiport rule, i.e., $a$ and $b$ are exchanged in membranes $i$ and
    $k$.

\item $i=k\neq j\neq l$: the \textit{presence-move rule} moves the symbol
    $b$ from membrane $j$ to $l$, provided that there is a symbol $a$ in
    membrane $i$.

\item $i=j\neq k\neq l$: the \textit{separation rule} sends $a$ and $b$
    from membrane $i$ to membranes $k$ and $l$, respectively.

\item $k=l\neq i\neq j$: the \textit{joining rule} brings $a$ and $b$
    together to membrane $i$.

\item $i=l\neq j\neq k$ or $i\neq j=k\neq l$: the \textit{chain rule} moves $%
a$ from membrane $i$ to membrane $k$ while $b$ is moved from membrane $j$
to membrane $i$, i.e., where $a$ previously has been.

\item $i\neq j\neq k\neq l$: the \textit{parallel-shift rule} moves $a$ and $%
b$ in independent membranes.
\end{enumerate}

A minimal interaction tissue P system may have rules of several types as
defined above. With respect to the computational power of such systems we
immediately see that when only antiport1 rules or only symport2 rules are used,
the number of objects in the system cannot be increased, hence, such systems
can generate only finite sets of natural numbers.
However, if we allow uniport rules (i.e., rules of the form %
$(a,i)\to (a,k)$ specifying that, whenever an object $a$ is present in cell
$i$, this may be moved to cell $k$), then minimal interaction tissue P systems
with symport2 and uniport rules or with antiport1 and uniport rules become
tissue P systems with minimal symport or minimal symport and antiport,
respectively.

By combining conditional-uniport-in rules and conditional-uniport-out rules,
computational completeness can be achieved by simulating a register machine.
The best known construction from~\cite{AARFSV-BSW07} is using 14 cells, but it
is very probably that this number can be decreased. A register machine may be
also simulated by using only the parallel-shift rule with~19
cells~\cite{Verlanetal08}. In all other cases, when only one of the types of
rules defined above is considered, it is not even clear whether infinite sets
of natural numbers can be generated.

Another interesting problem is to investigate how an interaction rule may be
simulated by some restricted variants. Such a study may lead to a formulation
of sufficient conditions on how combinations of variants of rules $(a,i) ( b,j)
\rightarrow ( a,k) ( b,l)$ may guarantee that the system can be realized by
using only specific restricted variants of rules in an equivalent minimal
interaction tissue P system. After that, a system satisfying sufficient
conditions of several restrictions may be automatically rewritten in terms of
any corresponding restricted variants. A list of such results can be found
in~\cite{Verlanetal08}.

\section{Number of Symbols}

Another complexity parameter that can be investigated is the number of objects
that can be used. The main results for P systems with antiport (and symport)
rules can be summarized in the following table:

%\begin{table}[hbt]
\begin{center}
%\caption{Families $NO_{m}P_{n}$}\vspace{2mm}
%\label{tableP}%
{\footnotesize
\begin{tabular}{c|cccccl}
\multicolumn{1}{c}{objects} &  &  &  &  & &  \\
5 & NRE &  &  &  &   &\\
4 & 2  & NRE &  &  &  & \\
3 & 1 & 2 & NRE &  &  & \\
2 & C & 1 & 2 & NRE &  & \\
1 & A & B & B  & B & B &\\\cline{2-6}
\multicolumn{1}{c}{}& 1 & 2 & 3 & 4 & $m\ (\ge5)$ & membranes \\
\end{tabular}%
}
\end{center}
%\end{table}

In the above table, the class of P systems indicated by A generates exactly
$NFIN$, the class indicated by B generates at least $NREG$, in the case of C at
least $NREG$ can be generated and at least $NFIN$ can be accepted, while a
class indicated by a number $d$ can simulate any $d$-register machine. The most
interesting questions still remaining open are to characterize the families
generated or accepted by P systems with only one symbol.

In the tissue case the situation changes as the additional links between every
cell and the environment permit to easily simulate a register
machine~\cite{AlhazovetalDLT05}. However, the definition used by the authors is
slightly different and it imposes a sequentiality for the communication between
two cells, i.e. if two rules that involve same two cells may be applied at the
same time, then only one of them will be chosen. The table below shows the
obtained results. In the table  $A$ indicates that the corresponding family
includes at least $NREG$, and $B$ indicates that the corresponding family can
generate more than $NFIN$.

%\begin{table}[htb]
%\caption{Families $NO_{m}tP_{n}$}
%\label{tabletP}%
\begin{tabular}{r|cccccccl}
\multicolumn{1}{c}{objects} &  &  &  &  &  &  &  &  \\
4 & $NREG$ & $NRE$ &  &  &  &  &  &  \\
3 & $NREG$ & $A$ & & & & & & \\
2 & $NREG$ & $A$ & $NRE$ & $NRE$ &$NRE$ & $NRE$ & $NRE$ & \\
1 & $NFIN$ & $B$ & $A$ & $A$ & $A$ & $A$ & $NRE$ & \\ \cline{2-8}
\multicolumn{1}{l}{} & 1 & 2 & 3 & 4& 5& 6& 7 & cells\\
\end{tabular}%
%\end{table}

\section{Number of Rules}

In this section we consider universal symport/antiport P systems having a small
number of rules. Such a bound can be obtained if we simulate a universal device
for which a bound on the number of rules is already known. Since P systems with
antiport and symport rules can easily simulate register machines, it is natural
to consider simulations of register machines having a small number of
instructions. An example of such a machine is the register machine $U_{32}$
described in \cite{Korec96}, which has $22$ instructions ($9$ increment and
$13$ decrement instructions). The table below summarizes the best results known
on this topic, showing the trade-off between the number of antiport rules and
their size:

%\begin{table}[htb]
%\caption{P systems with small numbers of antiport rules}
%\label{tabletPsmallnumbers}
\begin{center}
\begin{tabular}{|r|c|c|c|c|c|c|}\hline
number of rules & 73 & 56 & 47 & 43 & 30 & 23\\ \hline
size of rules & 3 & 5 & 6 & 7 & 11 & 19\\ \hline%
\end{tabular}%
\end{center}
%\end{table}

The results for columns 1, 4 and 5 were established
in~\cite{CsuhajVarjuetal07}, while other results are taken
from~\cite{AlhazovVerlan08}. The last column in this table is particularly
interesting, because the register machine $U_{32}$ which was the starting point
of the construction uses 25 computational branches.

\section{Conclusions}

Symport/antiport P systems were heavily investigated (there are more than 60
articles on this topic) and a lot of results about them are known, in
particular, about systems having low complexity parameters. This information
combined with their simple construction makes them an ideal object to be used
in universality proofs where they can replace register machines, in particular
for parallel computing devices. They are particularly well suited as a
simulated device for different classes of P systems which permits to obtain
different descriptional complexity improvements.

Even if there are a lot of results on P systems with symport/antiport, there
remain a lot of open questions; we would like to highlight the importance of
the investigation of generalized minimal communication models as this can show
new communication strategies that can be further used in other variants of P
systems. Another important topic is the number of rules of universal antiport P
systems with one membrane. This is especially interesting because such systems
directly correspond to maximally parallel multiset rewriting systems (MPMRS),
see~\cite{AlhazovVerlan08} for a formal definition of MPMRS. Since almost all
types of object-based P systems can be represented in terms of MPMRS, this will
give a lower bound on the number of rules needed for an universal P system.

\paragraph*{Acknowledgments} The authors
gratefully acknowledge support by the Science and Technology Center in Ukraine,
project 4032.

\bibliographystyle{eptcs}

\end{document}